\documentclass[12pt,aps]{revtex4}

\usepackage{bbm}
\usepackage{graphicx}
\usepackage{amssymb}
\usepackage{amsmath}

\newcommand{\nc}{\newcommand}
\nc{\be}{\begin{equation}}
\nc{\ee}{\end{equation}}
\nc{\bea}{\begin{eqnarray}}
\nc{\eea}{\end{eqnarray}}
\nc{\nn}{\nonumber}
     
\nc{\lp}{\left(}
\nc{\rp}{\right)}

\nc{\tr}{\textrm{tr}}

\begin{document}

\title{Triplet Leptogenesis in Left-Right Symmetric Seesaw Models}

\author{Tomas H\"allgren}
\email[]{tomashal@kth.se}

\author{Thomas Konstandin}
\email[]{konstand@kth.se}

\author{Tommy Ohlsson}
\email[]{tommy@theophys.kth.se}

\affiliation{Department of Theoretical Physics, Royal Institute of Technology (KTH), 
        AlbaNova University Center, Roslagstullsbacken 21, 106 91
        Stockholm, Sweden}


\begin{abstract}
We discuss scalar triplet leptogenesis in a specific left-right
symmetric seesaw model. We show that the Majorana phases that are
present in the model can be effectively used to saturate the existing
upper limit on the CP-asymmetry of the triplets. We solve the relevant
Boltzmann equations and analyze the viability of triplet
leptogenesis. It is known for this kind of scenario that the
efficiency of leptogenesis is maximal if there exists a hierarchy
between the branching ratios of the triplet decays into leptons and
Higgs particles. We show that triplet leptogenesis typically
favors branching ratios with not too strong hierarchies, since maximal
efficiency can only be obtained at the expense of suppressed
CP-asymmetries.

\end{abstract}

\maketitle

%
%

\section{Introduction}

Today, one of the major challenges of particle physics and cosmology
is to find a convincing model, which accounts for the observed baryon
asymmetry of the Universe. The baryon-to-photon ratio is observed to
be~\cite{Spergel:2003cb}
\begin{equation}
\eta_{B}=\frac{n_B-n_{{\bar B}}}{n_{\gamma}}=(6.1\pm 0.2)\times 10^{-10}.
\end{equation}
The introduction of superheavy right-handed neutrinos may account for
the baryon asymmetry by the baryogenesis via leptogenesis
mechanism~\cite{Fukugita:1986hr}, whereby the decays of the heavy
neutrinos induce a lepton asymmetry, which is then partly converted
into a baryon asymmetry by sphaleron processes.  This mechanism has
the advantage of simultaneously accounting for the smallness of light
neutrino masses via the seesaw
mechanism~\cite{Minkowski:1977sc,Gell-Mann:1980vs,Yanagida:1979as,
Glashow:1979nm,Mohapatra:1979ia, Magg:1980ut, Lazarides:1980nt,
Schechter:1980gr,Mohapatra:1980yp}.

Recently, it was shown in refs.~\cite{Akhmedov:2005np,Akhmedov:2006de}
that in left-right symmetric seesaw models, there exists an eight-fold
degeneracy in the right-handed neutrino sector, i.e., for a given
low-energy neutrino phenomenology, the seesaw formula can be inverted
to yield eight possible solutions for the triplet Yukawa coupling
matrices. It was then demonstrated in
refs.~\cite{Akhmedov:2006yp,Hosteins:2006ja} that by using fine-tuning
and viability of leptogenesis as criteria, it is possible to
discriminate among the different solutions. Another nice feature of
this model~\cite{Akhmedov:2006yp} is that the additional Majorana
phases can be used to easily saturate the existing upper bounds on the
CP-asymmetry of the heavy right-handed neutrinos derived in
ref.~\cite{Antusch:2004xy}. Besides, left-right symmetric models do
not contain many more parameters than the minimal seesaw type I model,
but can be naturally embedded into grand unified theories (GUTs),
since a hierarchical Yukawa coupling does not necessarily lead to a
fine-tuning problem in the right-handed neutrino
sector~\cite{Akhmedov:2006yp}. In these studies, it was assumed that
leptogenesis is solely driven by the decay of the lightest
right-handed neutrino.

In the present paper, we study the impact of leptogenesis induced by
the decay of the scalar $SU(2)_L$ triplet, which appears naturally in
left-right symmetric seesaw models. The common lore about triplet
leptogenesis is that it is far less efficient than leptogenesis driven
by right-handed neutrino decay, since the triplet can annihilate via
gauge interactions. This picture was challenged in
ref.~\cite{Hambye:2005tk}, where the full set of Boltzmann equations
for this model were investigated and it was shown that, due to the
interplay of the two possible decay channels of the triplet, the
efficiency $\eta$ can be quasi-maximal. In addition, an upper bound
on the CP-asymmetry as well as a lower limit on the triplet mass were
derived in this work in a model-independent seesaw type I+II setting
assuming hierarchical light neutrino masses:
\begin{equation}\label{eq:massbound0}
M_T >2.8\times 10^{10}~{\rm GeV}\,(\tilde m_T = 0.001~{\rm
eV}),\quad\quad M_T>1.3\times 10^{11}~{\rm GeV}
\,(\tilde m_T = 0.05~{\rm eV}). 
\end{equation}

In the present work, we will demonstrate that in the left-right
symmetric seesaw model the bound on the CP-asymmetry of the triplet
decay rate given in ref.~\cite{Hambye:2005tk} can be approximately
saturated.  This is due to the presence of additional Majorana phases
that, compared to the pure type I seesaw model, provides an additional
source of CP violation. Besides, we solve the relevant Boltzmann
equations and determine the produced baryon asymmetry. We discuss in
detail the occurrence of quasi-maximal efficiency of leptogenesis in
the case of a hierarchy in the branching ratios of triplet decays into
leptons and Higgs particles, $B_L \ll B_H$ or $B_L \gg B_H$. In
principle, this effect arises because in the decay channel with small
branching ratio washout is negligible. This implies that the large
efficiency is obtained at the expense of a suppression in the
CP-asymmetry that is normalized to the total decay width. These two
effects compete with each other and we demonstrate that leptogenesis
in fact favors the region in parameter space in which the rate of
triplet decays and annihilations are comparable, $\gamma_A\approx
\gamma_D$ at $ M_T \approx T$, and also the branching ratios are of
similar size, $B_L \approx B_H$.

The paper is organized as follows. In section~\ref{sec_model} we
present the model and set up the notation. In section~\ref{sec_lep}
we discuss triplet leptogenesis, and in section~\ref{sec_concl} we
draw our conclusions.

\section{Left-right symmetric model\label{sec_model}}

In this section, the left-right symmetric model of
refs.~\cite{Hambye:2005tk,Akhmedov:2006yp} is presented. The
Lagrangian contains the following relevant terms
\bea
\label{Lagrange_LR}
\mathcal{L} &\supset&
-\frac12 M_T^2{\rm tr}\left(T_{L}^{\dagger}T_{L}\right)
+ \frac12 \kappa_{ij}{\rm tr}\left(T_{R}^{\dagger}\Phi_{i}T_{L}\Phi_{j}^{\dagger}\right) \nn \\
&&- \frac12 f^{\alpha\beta}L_{\alpha}^{T}C{\rm i}\tau_{2}T_{L}L_{\beta} 
- \frac12 f^{\alpha\beta}R_{\alpha}^{T}C{\rm i}\tau_{2}T_{R}R_{\beta} \nn \\
&&
+ y_i^{\alpha\beta} {\bar R}_{\alpha} \Phi_{i} L_{\beta}   + {\rm h.c.},
\eea
where $L$ and $R$ denote the left- and right-handed leptons,
respectively. The Yukawa coupling is assumed to be complex symmetric,
i.e., $y_i=y_i^T$, what holds true in certain left-right symmetric
GUTs. The Higgs bidoublet fields $\Phi_{i}$ are given by
\begin{equation}
\Phi_{1}=\begin{pmatrix} \Phi_{1}^{0} && \Phi_{1}^{+} \\
\Phi_{2}^{-} && \Phi_{2}^{0}
\end{pmatrix}, \quad
\quad\quad \Phi_{2}=\tau_{2}\Phi_{1}^{*}\tau_2,
\end{equation}
and the triplets $T_{L/R}$ can be written as
\begin{equation}
T_{L/R}=\left(\begin{matrix} T^{+}/\sqrt{2} && T^{++}\\
T^{0} && -T^{+}/\sqrt{2}\end{matrix}\right)_{L/R}.
\end{equation}
By spontaneous symmetry breaking, the Higgs fields acquire vacuum
expectation values (vevs) that are related by
\bea
\label{rel_vevs}
v_L &=& \frac{v_R}{2 M_T^2} \kappa v^2,  \\
\kappa v^2 &=& \kappa_{11} v_2 v^*_1 + \kappa_{12} v_2^2 + 
\kappa_{21} v_1^{*2} + \kappa_{22} v_1^* v_2,
\eea
where $v_{L/R}$ denote the vevs of the $SU(2)_{L/R}$ scalar triplets
$T_{L/R}$, $v^2 = |v_1^2| + |v_2^2|\simeq(174 \textrm{ GeV})^2$,
and $v_i=\left< \Phi_i^0 \right>$ are the vevs of the bidoublet.

The resulting neutrino masses lead to the following seesaw
formula in the left-right symmetric model
\begin{equation}\label{eq:seesaw}
m_{\nu}=m_T+m_H = f v_L - \frac{v^2}{v_R}yf^{-1}y.
\end{equation}
The contribution $m_H$ from the heavy right-handed neutrinos is hereby
usually called the type I term, while the contribution $m_T$ from the
triplet is called the type II term. This equation can be simplified by
using the fact that the combination $m_T=f v_{L}$ depends on the
parameter
\be
\mu = \frac{v_{R}}{v_L v^2}
\ee
only and not on $v_L$ and $v_R$ separately. Given a light neutrino
mass matrix $m_{\nu}$ and Yukawa coupling matrix $y$, there exists for
$n$ flavors $2^n$ solutions for the triplet Yukawa coupling matrix $f$
as shown in refs.~\cite{Akhmedov:2005np,Akhmedov:2006de}. In the case
of one flavor, the inversion becomes particularly simple and one finds
the following two solutions
\begin{equation}
\label{f1flavor}
f_{\pm} =
\frac{m_{\nu}}{2v_{L}}\pm\sqrt{\frac{m_{\nu}^2}{4v_{L}^2}
+\frac{v^{2}}{v_{L}v_R}y^2}.
\end{equation}
An analytical expression for the inversion of the seesaw relation in
the three-flavor case is also given in
refs.~\cite{Akhmedov:2005np,Akhmedov:2006de}. The properties of these
solutions have been studied in
refs.~\cite{Akhmedov:2006de,Akhmedov:2006yp,Hosteins:2006ja}, and in
particular, the eigenvalues and mixing properties were
investigated. As in these works, we choose the Yukawa coupling matrix
$y$ equal to the up-type quark Yukawa coupling matrix, which can be
motivated by GUTs~\cite{Pati:1973uk,Georgi:1974my,Fritzsch:1974nn}. To
be specific, we implement the relation $y=y_\textrm{up}$ at the GUT
scale and in the flavor basis. The CKM matrix is completely attributed
to the up-quark Yukawa matrix. We utilize the best-fit values for the
parameters of the neutrino mass matrix as given in
refs.~\cite{Maltoni:2004ei, Strumia:2005tc, Fogli:2006qg}. It is known
that depending on the spectrum of the light neutrinos, running effects
can also be sizable in the neutrino sector. In the present study we
neglect these effects, since the specific values of the mixing angles
will not strongly influence our results concerning leptogenesis. This
is due to the fact that our main source of CP violation stems from the
additional Majorana phases and not from the Dirac phases in the mixing
matrices. As additional parameters of the model enter then the five
Majorana phases, the mass of the lightest neutrino $m_0$, the
hierarchy (normal/inverted) of the light neutrinos, and the ratio
$v_R/v_L$.

\begin{figure}
\begin{center}
\includegraphics*[width=\textwidth]{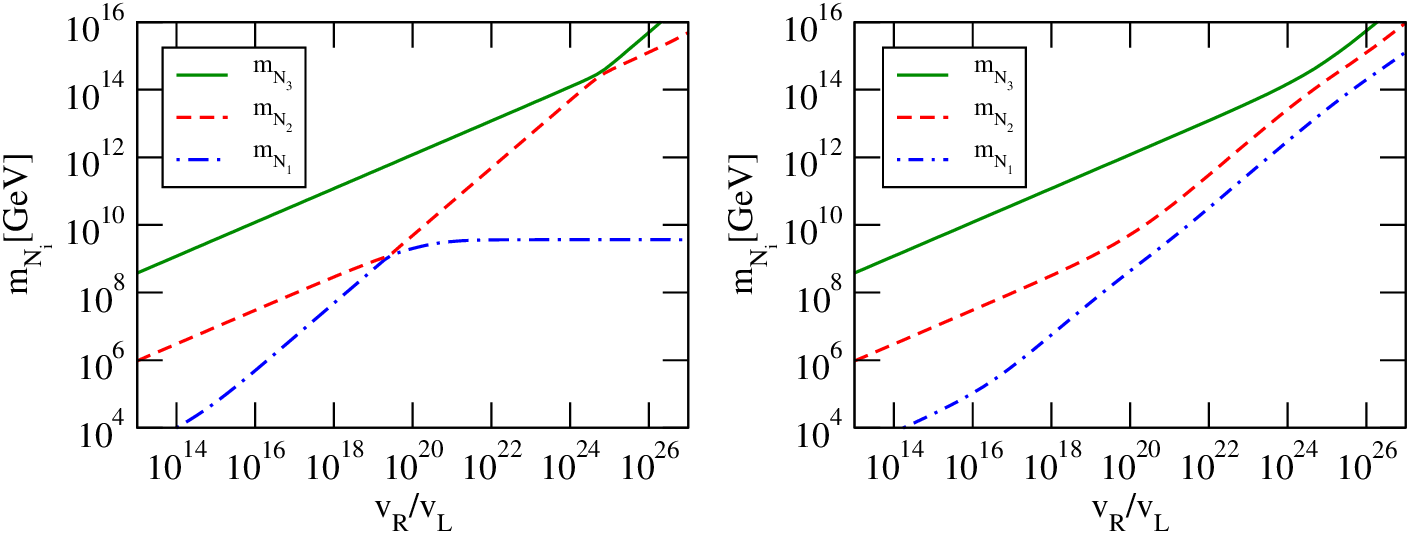}
\end{center}
\vskip -1cm
\caption{\small
Right-handed neutrino masses as functions of $v_R/v_L$ for the two
solutions '$+-+$' (left plot) and '$+++$' (right plot). }
\label{fig_solutions}
\end{figure}

Before we discuss triplet leptogenesis, we remark on some of the
general properties of the solutions for the right-handed neutrino
masses. In the regime of small $v_R/v_L$, both contributions in
eq.~(\ref{eq:seesaw}) are large and the light neutrino mass matrix
results from a precise cancellation of both contributions.  This
requires a large amount of fine-tuning, and hence constitutes a
disfavored region in parameter space. In the regime of large
$v_R/v_L$, the solutions can be labeled according to whether the
eigenvalues are type I ('$-$') or type II ('$+$') dominated. This
corresponds to the choice of the different signs appearing in
eq.~(\ref{f1flavor}) and results in the eight different solutions in
the case of three flavors. If an eigenvalue is type I dominated, it
approaches a constant value for large $v_R/v_L$, which is given by the
corresponding eigenvalue of the Yukawa coupling matrix $y$. Thus, the
smallness of the up-quark Yukawa coupling leads to the fact that the
four solutions of type '$\pm\pm-$' predict a very light right-handed
neutrino with a mass below $10^6$ GeV. These solutions are disfavored
from the point of view of leptogenesis. For the two other solutions of
type '$\pm-+$', the lightest mass instead approaches $10^9$ GeV in the
large $v_R/v_L$ limit, which may induce viable leptogenesis, while
being consistent with the bounds on the reheating temperature from
gravitino physics as described in detail in
ref.~\cite{Akhmedov:2006yp}. The remaining two solutions predict
larger values for the right-handed neutrino masses in the large
$v_R/v_L$ regime. These facts are demonstrated in
Fig.~\ref{fig_solutions}, where the masses of the right-handed
neutrinos, $m_N \propto f\, v_R$, for the two solutions '$+-+$' and
'$+++$' are displayed.

\section{Triplet leptogenesis}\label{sec_lep}

In the traditional framework of leptogenesis, the CP-asymmetry is
induced by the decay of the lightest right-handed neutrino. Here we
give a brief review of the case when instead the triplet is the source
of leptogenesis. In this case, the CP-asymmetry has been studied in
refs.~\cite{Sarkar:1998cf, Lazarides:1998jt, Hambye:2003ka}. The full
set of Boltzmann equations were derived and solved in
ref.~\cite{Hambye:2005tk}. Generally, one would expect that the
produced baryon asymmetry is suppressed, since the gauge scattering
keeps the triplets close to thermal equilibrium, preventing sufficient
out-of-equilibrium decays.  However, it was shown in
ref.~\cite{Hambye:2005tk} that, due to an interplay of the two
possible decay channels of the triplet, a quasi maximal efficiency can
be achieved.

The Lagrangian in ref.~\cite{Hambye:2005tk} contains the following
relevant terms
\begin{equation}
\mathcal{L}\supset- \frac12 M_T^2{\rm
tr}\left(T_{L}^{\dagger}T_{L}\right)
-\frac{1}{2} \lambda^{\alpha\beta}_{L}L_{\alpha}^{T}C{\rm
i}\tau_{2}T_{L}L_{\beta} + \frac{1}{2}\lambda_H
M_T H^{T}{\rm i}\tau_{2}T_{L}^\dagger H+{\rm h.c.}
\end{equation}
Comparing this with the Lagrangian in eq.~(\ref{Lagrange_LR}), one can
make the following two identifications~\cite{Hambye:2003ka}
\begin{equation}
\label{matching}
\lambda_H M_T = \kappa v_{R}, \quad
\lambda_L = f.
\end{equation}
Notice that perturbative unitarity in the left-right symmetric model
implies $\kappa\lesssim1$ instead of $\lambda_H \lesssim1$ as used in
ref.~\cite{Hambye:2005tk}. The former constraint is less severe, since
$f\lesssim 1$, and hence, $v_R \gtrsim m_{N_3} \gtrsim M_T$ in the
parameter region that is interesting for triplet leptogenesis.

\subsection{CP-asymmetry}

The CP-asymmetry is due to the decay of the triplet via the two
possible channels $T_L\rightarrow HH$ and $T_L\rightarrow
\bar L\bar L$, where $H$ and $L$ denote Higgs and lepton doublets,
respectively. We follow the notation of ref.~\cite{Hambye:2005tk} and
denote the branching ratio into Higgs bosons by $B_H$ and the
branching ratio into leptons by $B_L$, and assume that these are the
only possible decay channels, i.e., $B_L+B_H=1$. At tree level, the
decay rates are given by~\cite{Hambye:2005tk}
\bea
\label{decay_width}
\Gamma(T\to \bar L\bar L)&=&
B_L \Gamma_T=\frac{M_T}{16 \pi} \tr ( \lambda_L \lambda_L^\dagger ),  \\
\Gamma(T\to HH)&=&B_H \Gamma_T=\frac{M_T}{16 \pi} \lambda_H \lambda_H^\dagger.
\eea
If the triplet is lighter than the right-handed neutrinos, the
CP-asymmetry takes the form~\cite{Hambye:2005tk}
\begin{equation}\label{eq:epsilon}
\epsilon_L = \frac{M_T}{4\pi v^2}\sqrt{B_L B_H}\frac{{\rm
 Im}\left[{\rm
 tr}\left(m_T^{\dagger}m_{H}\right)\right]}{\tilde{m}_T}.
\end{equation}
In this expression, $m_T (m_H)$ denotes the triplet (right-handed
neutrino) contribution to the light neutrino mass as given in
eq.~(\ref{eq:seesaw}). In addition, we have introduced the parameter
\begin{equation}
\tilde{m}_T = \sqrt{{\rm tr}\left(m_T^{\dagger}m_T\right)}.
\end{equation}
Notice that using eqs.~(\ref{rel_vevs}) and (\ref{matching}) $m_T$ can
be recast as
\be
m_T = f \, v_L = \lambda_L \lambda_H \frac{v^2}{2M_T},
\ee
such that
\be
\Gamma_T = \frac{\tilde{m}_T}{\sqrt{B_L \, B_H}} \frac{M_T^2}{8 \pi v^2}.
\ee
An upper bound on the CP-asymmetry was also derived in
ref.~\cite{Hambye:2005tk} and found to be
\begin{equation} \label{eq:epsilonmax}
|\epsilon_L|\leq \frac{M_T}{4\pi v^2}\sqrt{B_L B_H\sum_i m_{\nu_i}^2}
=: \epsilon_{L,\textrm{max}},
\end{equation} 
where $m_{\nu_i}$ denote the light neutrino masses.

In our specific model, we use the following input parameters. As
mentioned in the last section, a reasonable choice for the Yukawa
coupling matrix $y$, motivated by GUTs, is the up-quark Yukawa
coupling. In addition, the Majorana phases, the ratio $v_R/v_L$, the
lightest neutrino mass $m_0$, and the hierarchy of the light neutrinos
have to be specified. Using these parameters, the seesaw formula in
eq.~(\ref{eq:seesaw}) can be used to determine $f v_L$, and thus, the
masses of the right-handed neutrinos and the parameter $\tilde m_T$.
The CP-asymmetry $\epsilon_L$ depends in addition on $B_L$ and $B_H$,
but only through the explicit factor in eq.~(\ref{eq:epsilon}).

The values of $B_L$ and $B_H$ depend in the case of fixed $v_R/v_L$ on
the choice of $v_R$ or $v_L$ according to
\be
\label{BLBHadjust}
\frac{B_L}{B_H} = 
\frac{\tr (\lambda_L \lambda_L^\dagger)}{\lambda_H \lambda_H^\dagger }
= \frac{\sum_i m^2_{N_i} }{4 v_R^2 v_L^2 M_T^2} v^4.
\ee
In the following, we plot the quantities $\tilde m_T$ and
$\epsilon_L/\epsilon_{L,\textrm{max}}$ that are both independent from the
choice of $B_H$, $B_L$, and $M_T$.
\begin{figure}
\begin{center}
\includegraphics*[width=\textwidth]{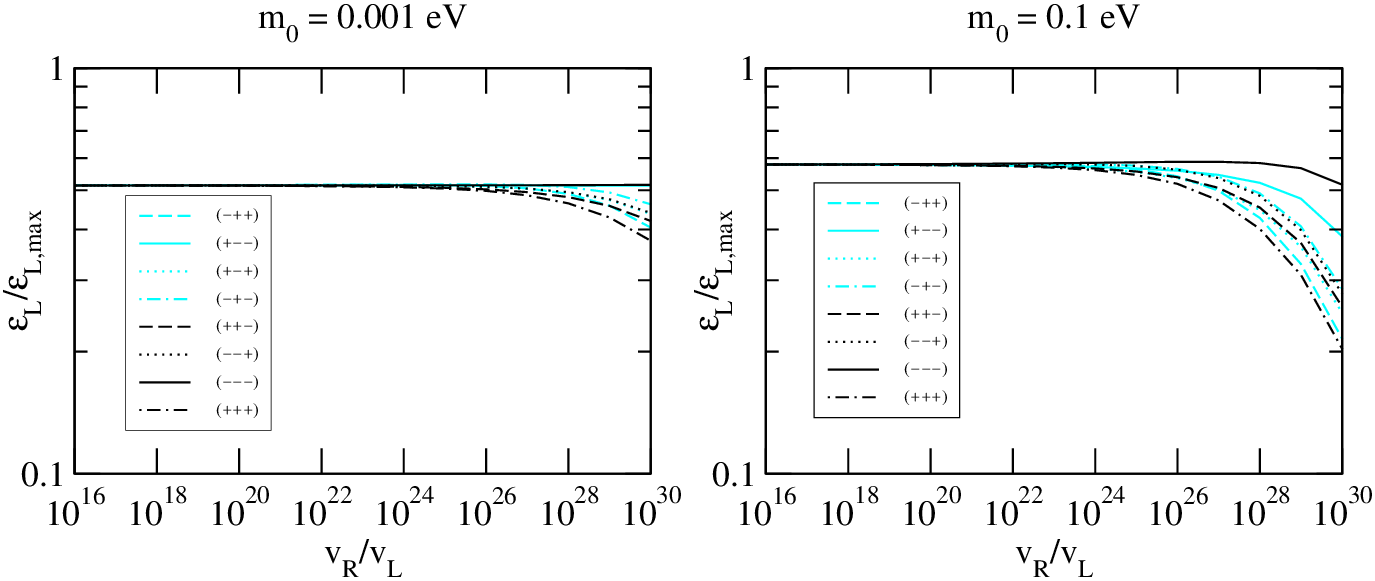}
\end{center}
\vskip-1cm
\caption{ \small
The parameter $\epsilon_L/\epsilon_{L,\textrm{max}}$ for all eight
solutions as a function of $v_{R}/v_{L}$. Inverted hierarchy and
$m_{0}=0.001$ eV ($m_{0}=0.1$ eV) in the left (right) plot. }
\label{fig:epsilonppp}
\end{figure}
\begin{figure}
\begin{center}
\includegraphics*[width=\textwidth]{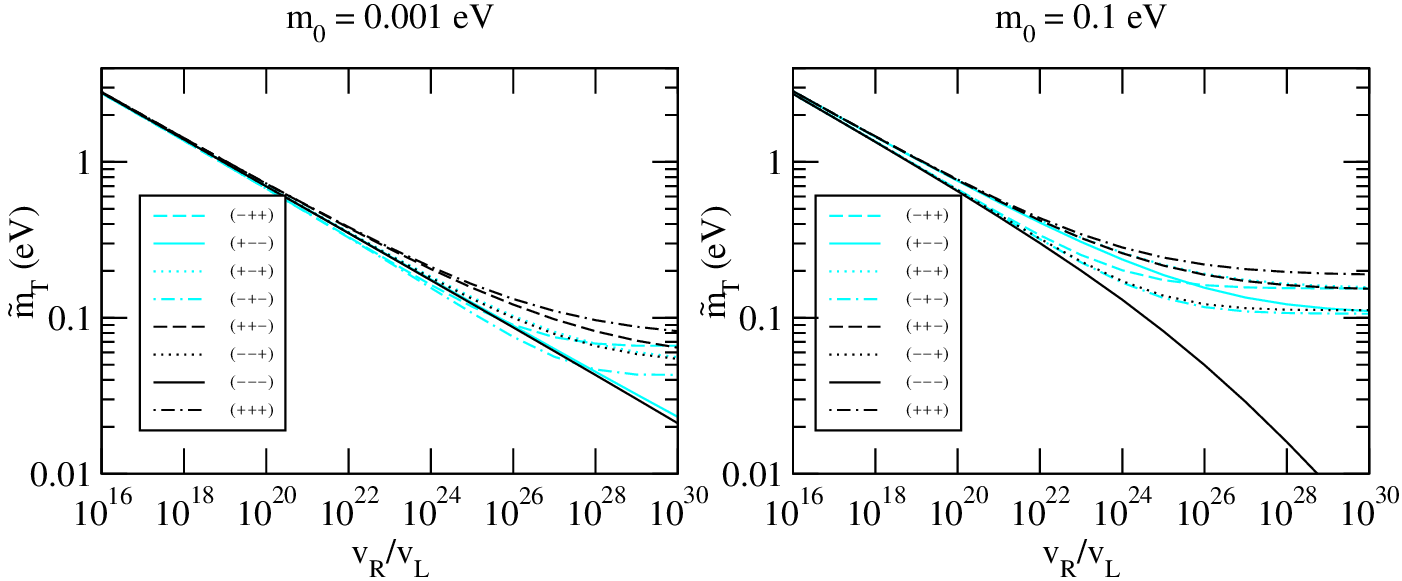}
\end{center}
\vskip-1cm
\caption{
\small \label{fig:mtildep001}
The parameter $\tilde{m}_T$ in units of eV for all eight
solutions as a function of $v_{R}/v_{L}$. Inverted hierarchy and
$m_{0}=0.001$ eV ($m_{0}=0.1$ eV) in the left (right) plot.  }
\end{figure}
Figure~\ref{fig:epsilonppp} shows that it is approximately possible to
saturate the bound on the CP-asymmetry stated in
eq.~(\ref{eq:epsilonmax}) for a certain choice of the Majorana
phases. Additionally, $\epsilon_L/\epsilon_{L,\textrm{max}}$ does not
depend on $v_R/v_L$ up to $v_R/v_L \approx 10^{28}$. On the other
hand, $\tilde m_T$ generally decreases for increasing $v_R/v_L$, as
shown in Fig.~\ref{fig:mtildep001}.


Let us summarize our findings in the left-right symmetric model
concerning triplet leptogenesis: One main parameter in the model is
the ratio $v_R/v_L$ that also enters into the masses of the
right-handed neutrinos. Efficient triplet leptogenesis requires the
lightest right-handed neutrino to be heavier than the triplet,
otherwise inverse decays into the neutrino tend to erase every lepton
asymmetry produced by triplet decays. In the light of the constraint
in eq.~(\ref{eq:massbound0}) and Fig.~\ref{fig_solutions}, this is
only the case for the two solution of type '$\pm++$' and $v_R/v_L
\gtrsim 10^{22}$. The CP-asymmetry of the triplet does not vary
strongly as long as $v_R/v_L\lesssim 10^{28}$ and for an appropriate
choice for the Majorana phases, the upper bound given in
eq.~(\ref{eq:epsilonmax}) can be approximately saturated. This upper
bound is proportional to $\sqrt{\sum_i m_{\nu_i}^2}$ such that large
values of $m_0$ seem to be opportune for triplet leptogenesis. On the
other hand, the effective mass parameter $\tilde m_T$ that will enter
in the Boltzmann equations seems to be constrained by $\tilde m_T
\gtrsim m_0$ such that larger values of $m_0$ might lead to larger
washout. This question will be discussed in the next section. The
value of $\tilde m_T$ can in a certain range be chosen by specifying
the ratio $v_R/v_L$. On the other hand, the parameters $B_L$ and $B_H$
can be adjusted by changing $v_R$ or $v_L$ for fixed $v_R/v_L$ according
to eq.~(\ref{BLBHadjust}). Hence, we will treat $B_L$ ($B_H$), $M_T$,
and $\tilde m_T$ as free parameters in the following discussion of the
transport equations.

\subsection{Boltzmann equations}

The complete set of Boltzmann equations has been derived in
ref.~\cite{Hambye:2005tk} and we will discuss some qualitative results
in the following. First, since triplets, different from Majorana
neutrinos, are not self-conjugated fields, the evolution of the number
asymmetry $n_{T}- n_{\bar{T}}$ is governed by an additional equation.
Besides, the system of equations describes the evolution of the number
density of the Higgs particles $n_H$. The rates that are most
important for determining the efficiency $\eta$ are the gauge
scatterings $\gamma_A$ and the triplet decays into Higgs bosons
($\gamma_H = B_H \, \gamma_D$) and leptons ($\gamma_L = B_L
\gamma_D$). If one of the decay rates of the triplet ($\gamma_H$ or
$\gamma_L$) is larger than the gauge scatterings, the triplets decay
before annihilating, thus making the gauge scatterings
inefficient. This would also imply that the triplet decay rate is
larger than the expansion rate, which would therefore seem to make
leptogenesis impossible. However, if the second decay rate is smaller
than the expansion rate, a lepton asymmetry can be induced. It is this
interplay of $\gamma_L$, $\gamma_H$, and $\gamma_A$, which can result
in a quasi-maximal efficiency in triplet leptogenesis, as demonstrated
in ref.~\cite{Hambye:2005tk}. On the other hand, a hierarchy between
$B_L$ and $B_H$ leads to a suppression in the CP-asymmetry as can be
seen in eq.~(\ref{eq:epsilonmax}) such that it is not guaranteed that
the produced baryon asymmetry is maximal in this region of the
parameter space.

Our results are obtained by solving a simplified set of transport
equations, where we include the $\Delta T=2$ scatterings as given in
ref.~\cite{Hambye:2005tk}, but neglect the $\Delta L=2$ scatterings
that are minute in most cases. The corresponding set of Boltzmann
equations are ($z=M_T/T$)
\bea
\label{eq:Boltzmann}
\label{B1}
sH z \frac{d \Sigma_T}{dz} &=& 
- \gamma_D \lp \frac{\Sigma_T}{\Sigma_T^{eq}} -1 \rp 
- 2 \gamma_A \lp \frac{\Sigma^2_T}{\Sigma_T^{2\,eq}} -1 \rp ,  \\  
\label{B2}
sH z \frac{d \Delta_L}{dz} &=& \epsilon_L \gamma_D 
 \lp \frac{\Sigma_T}{\Sigma_T^{eq}} -1 \rp 
- 2 \gamma_D B_L \lp \frac{\Delta_L}{Y_L^{eq}} + \frac{\Delta_T}{ \Sigma_T^{eq}}\rp,   \\  
\label{B3}
sH z \frac{d \Delta_H}{dz} &=& \epsilon_L \gamma_D 
 \lp \frac{\Sigma_T}{\Sigma_T^{eq}} -1 \rp 
- 2 \gamma_D B_H \lp \frac{\Delta_H}{Y_H^{eq}} - \frac{\Delta_T}{ \Sigma_T^{eq}}\rp,   \\  
\label{B4}
sH z \frac{d \Delta_T}{dz} &=&   
- \gamma_D  \lp \frac{\Delta_T}{ \Sigma_T^{eq}} 
+ B_L \frac{\Delta_L}{Y_L^{eq}} - B_H \frac{\Delta_H}{Y_H^{eq}} \rp,
\eea
where $\Sigma_T$, $\Delta_X$, and $Y_X$ denote particle numbers
normalized to the entropy
\be
\Sigma_T = (n_T + n_{\bar T})/s, \quad
\Delta_X = (n_X - n_{\bar X})/s,  \quad
Y_X = n_X/s, 
\ee
that are in equilibrium given by 
\be
s = g_* \frac{2 \pi^2}{45} T^3, \quad
n_\gamma = 2 \frac{\zeta(3)}{\pi^2} T^3, \quad
n^{eq}_L = \frac{27}{4} n_\gamma, \quad
n^{eq}_H = 2 n_\gamma , \quad 
n^{eq}_T = \frac{3}{4} \, n_\gamma \, z^2 K_2( z ).
\ee
The Hubble parameter is $H \simeq 1.66 \sqrt{g_*} T^2/M_{Pl}$ and the
rate of triplet decays and inverse decays is given by
\be
\gamma_D = s \Gamma_T \Sigma^{eq}_T K_1(z) / K_2(z),
\ee
where $\Gamma_T$ is the decay width as given in
eq.~(\ref{decay_width}). The annihilation rate $\gamma_A$ turns out to
be
\bea
\gamma_A(z) &=&\frac{T M_T^3}{64 \pi^4} \int_4^\infty dx \sqrt{x} 
K_1(z\sqrt{x}) \hat \sigma(x),  \\
\hat \sigma(x) &=& \frac{50 g_2^4 + 41 g_Y^2}{16 \pi} r^3 
	 + \frac{r}{2\pi} \left[ g_2^4 (10+68/x) + g_Y^4 (1 + 4/x) \right] \nn \\
	&& + \frac{1}{2\pi x^2} \log{\frac{1+r}{1-r}} 
\left[ g_2^4 (48 x-48) + g_Y^4 (12x -24) \right]
\eea
with $r=\sqrt{1-4/x}$ and $g_2$ ($g_Y$) denotes the $SU(2)_L$ ($U(1)_Y$)
gauge coupling of the Standard Model. The efficiency $\eta$ is
defined by
\be
 \Delta_L  = \epsilon_L \eta \left.
\Sigma_T \right|_{T \gg M_T},
\ee
such that the baryon-to-photon ratio is at late times given by
\be
\eta_B = \frac{n_B}{n_\gamma} = -\frac{28}{79} \frac{s}{n_\gamma} \Delta_L 
\simeq -0.039 \, \epsilon_L \, \eta.
\ee

Before we present numerical results, we analyze the qualitative
behavior of the solutions to the transport equations in certain
limiting cases. First of all, notice that eq.~(\ref{B1}) that described
the dynamics of the total triplet number density decouples from the
other equations. An explicit solution is given by
\bea
\label{eq:source}
\mathcal{S}(z) &=& \frac{\gamma_D}{s H} \lp \frac{\Sigma_T}{\Sigma_T^{eq}} -1 \rp \\
&\approx& \frac{\gamma_D}{s H \Sigma_T^{eq}} \int^z \, dz' \, 
\frac{d \Sigma_T^{eq}(z') }{dz'} 
\exp \lp - \int_{z'}^z  \frac{d\tilde z}{\tilde z} 
\frac{\gamma_D + 4 \gamma_A}{s H \Sigma_T^{eq}} (\tilde z) \rp.
\eea
Here, the function $\mathcal{S}$ was defined that will act as a source
for the lepton and Higgs asymmetries. The magnitude of the source
depends on the two decay rates that are to high precision given by the
approximations
\bea
\label{eq:gammaapprox}
\frac{\gamma_D}{sH} &\simeq& \frac{1}{\sqrt{B_L B_H}} 
\frac{\tilde m_T}{ 10^{-3} \textrm{ eV }} z^4 K_1(z), \\
\frac{\gamma_A}{sH} &\simeq& \frac{10^{14} \textrm{ GeV }}{M_T}
z^4 K_2(2z) \, \lp \frac{1}{1+25z^2} \rp^{1/4}.
\eea

For small values of the triplet mass, $M_T \lesssim 10^{11}$ GeV, the
annihilation rate is strong enough to keep the triplets close to
equilibrium when they become non-relativistic. Hence, the source
increases linearly with the decay rate as long as $\gamma_D \ll
\gamma_A$, but becomes exponentially suppressed for $\gamma_D \gg
\gamma_A$. The branching ratio that maximizes the produced baryon
asymmetry depends crucially on the question if triplet decays are
faster than annihilation processes or vice versa, as will be discussed
in the following.

From the three equations that determine the dynamics of the lepton
number and Higgs asymmetries, eq.~(\ref{B4}) is redundant, since
$U(1)_Y$ conservation implies that
\be
2\Delta_T=\Delta_L-\Delta_H,
\ee
and the remaining two equations are then of the form
\bea
 z \frac{d \Delta_L}{dz} &=& \epsilon_L \mathcal{S}
- 2 B_L \frac{\gamma_D}{H s}
\lp \frac{\Delta_L}{Y_L^{eq}} + \frac{\Delta_L - \Delta_H}{2 \Sigma_T^{eq}}\rp,   \\  
 z \frac{d \Delta_H}{dz} &=& \epsilon_L \mathcal{S}
- 2 B_H \frac{\gamma_D}{H s}
\lp \frac{\Delta_H}{Y_H^{eq}} - \frac{\Delta_L - \Delta_H}{2 \Sigma_T^{eq}}\rp.  
\eea
Since $\Delta_T$ is proportional to $\Delta_L-\Delta_H$, the effect of
eq.~(\ref{B4}) is to distribute a generated asymmetry between the two
channels. This process is dominant at late times, when all triplets
decay and $\Sigma_T^{eq} \ll Y_{L/H}$, such that $\Delta_L=\Delta_H$
for $z \to \infty$. Thus, it is not important in which channel the
asymmetry was generated initially.

\begin{figure}
\begin{center}
\includegraphics*[width=\textwidth]{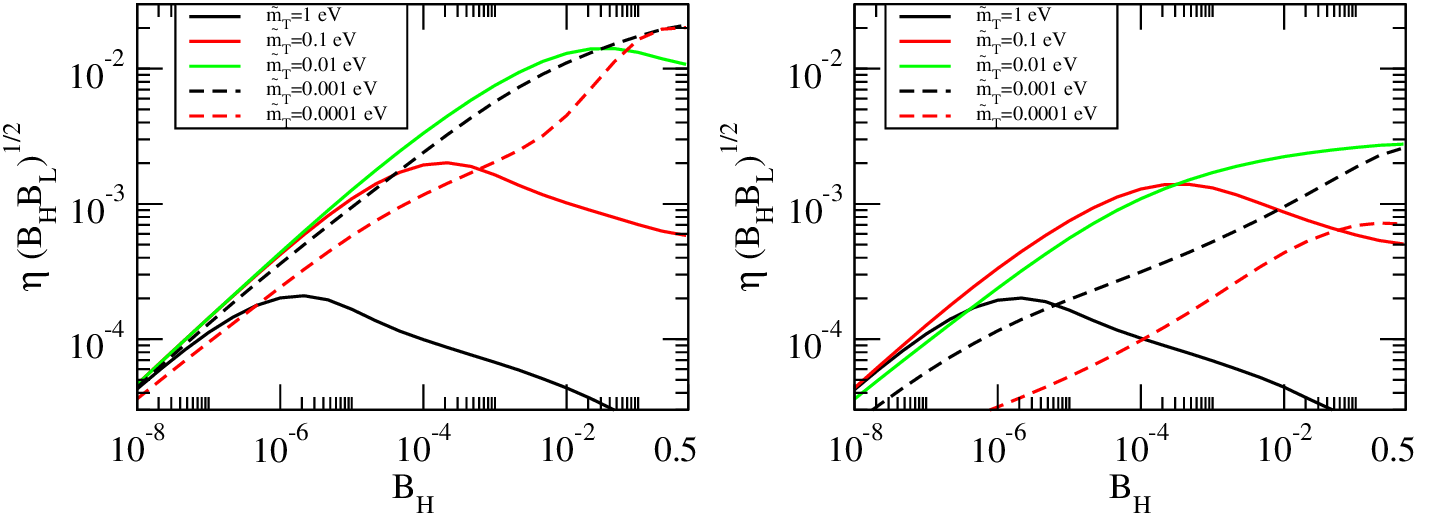}
\end{center}
\vskip-1cm
\caption{
\small \label{eta_plot}
The quantity $\eta \sqrt{B_L B_H}$ as a function of $B_H$ for $M_T = 2
\times 10^{12}$ GeV (left plot) and $M_T = 2 \times 10^{10}$ GeV 
(right plot) and five different values of the parameter $\tilde m_T$.
}
\end{figure}
\begin{figure}
\begin{center}
\includegraphics*[width=\textwidth]{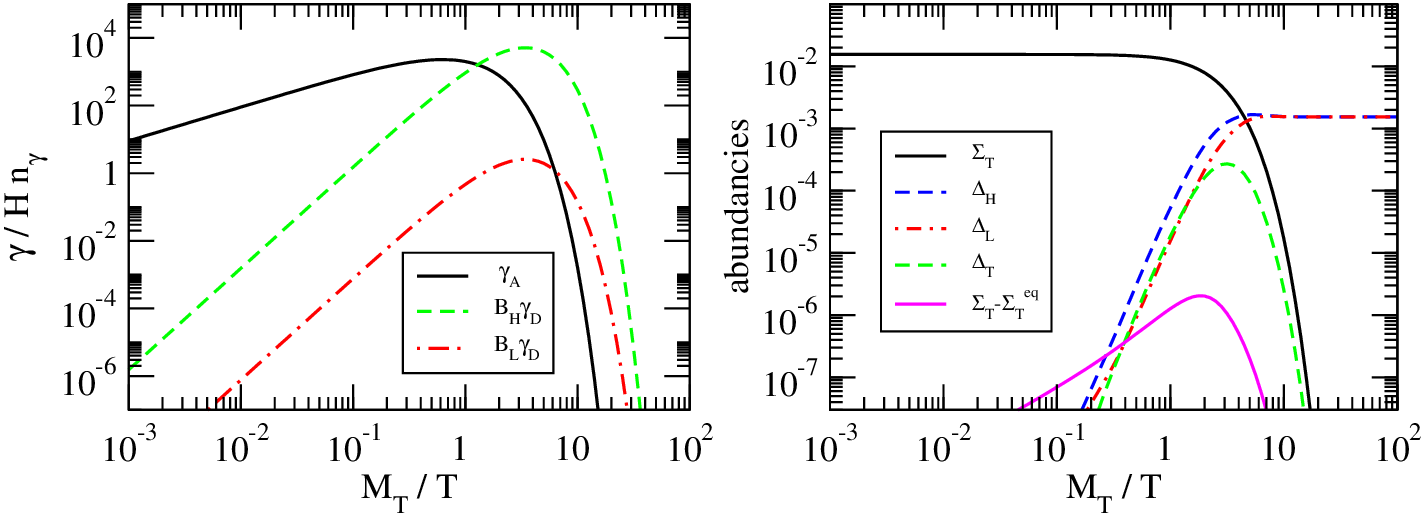}
\end{center}
\vskip -1cm
\caption{\small
The left plot shows the decay rates $\gamma_A$, $B_H \gamma_D$, and
$B_L \gamma_D$ in units of $H n_\gamma$. The right plot shows the
abundances $\Sigma_T$, $\Delta_L$, $\Delta_H$, $\Delta_T$, and
$\Sigma_T - \Sigma_T^{eq}$ for $\epsilon_L=1$. The used parameters
are $M_T=10^{11}$ GeV, $B_H=5 \times 10^{-4}$, and $\tilde m_T=0.05$
eV resulting in $\eta=0.1$.}
\label{fig_trans}
\end{figure}

First, we discuss the limit $\gamma_D \gg \gamma_A$ when annihilation
processes are negligible. If one of the branching ratios, e.g. $B_H$,
is much smaller than the other, washout is only operative in one
channel, while the other channel accumulates a significant
asymmetry. Nevertheless, this asymmetry is partially reduced by the
mixing with the triplet asymmetry $\Delta_T$ and finally equally
distributed between the two channels. In this regime, the efficiency
$\eta$ does not depend strongly on the branching ratios and, comparing
with eq.~(\ref{eq:epsilon}), one expects that the produced baryon
asymmetry is proportional to $\sqrt{B_L B_H} \approx \sqrt{B_H}$.  On
the other hand, for larger values of $B_H$ washout is significant and
the baryon asymmetry produced by the Higgs decay channel starts to be
exponentially suppressed for
\be
B_H \frac{\gamma_D}{H n_\gamma} \gtrsim 1,
\ee
such that according to eq.~(\ref{eq:gammaapprox}) the optimal choice
for $B_H$ (and likewise for $B_L$) scales as
\be
B_\textrm{max} \propto \tilde m_T^{-2}.
\ee
This is demonstrated in Fig.~\ref{eta_plot} that shows $\eta \sqrt{B_L
B_H}$ as a function of $B_H$ for $M_T=2\times10^{12}$~GeV and
$M_T=2\times10^{10}$~GeV and five different values of $\tilde m_T$.
Notice also that in this limit the annihilation rate and consequently
the triplet mass is not relevant for the efficiency~$\eta$.
Since the branching ratios $B_L$ and $B_H$ enter in the CP-asymmetry
only by the explicit factor in eq.~(\ref{eq:epsilon}), we show in
Fig.~\ref{eta_plot} the combination $\eta \sqrt{B_L B_H}$ that is
proportional to the baryon asymmetry $\eta_B$.  A specific example for
a solution of the transport equations is given in
Fig.~\ref{fig_trans}.

In the opposite limit, $\gamma_D \ll \gamma_A$, the source
$\mathcal{S}$ is proportional to $\gamma_D$, according to
eq.~(\ref{eq:source}). Washout is not relevant and the efficiency
$\eta$ is independent of $B_L$ and $B_H$ even for large branching
ratios. Hence, the predicted baryon asymmetry is maximal for the
choice $B_L\approx B_H
\approx 1/2$, since in this case the CP-asymmetry is maximal. This
behavior can be seen in Fig.~\ref{eta_plot}. Notice that in the left
plot the limit $\gamma_D \approx \gamma_A$ is barely reached.

With respect to the effective mass $\tilde m_T$, leptogenesis favors
the region in parameter space, where the decay rate and the
annihilation rate are comparable, $\gamma_D \approx \gamma_A$ at
$z\approx 1$.  In the case of a rather heavy triplet (left plot of
Fig.~\ref{eta_plot}), annihilations are suppressed such that the
situation is similar to leptogenesis driven by right-handed neutrino
decays. Leptogenesis is maximal when the decay rate is small, $\tilde
m_T \lesssim 0.01$ eV, and also the branching ratios are of similar
size, $B_L \approx B_H$.  In the case of a smaller triplet mass (right
plot of Fig.~\ref{eta_plot}), annihilations become important and
compete with the two decay processes. Still, a large hierarchy between
the branching ratios is not required for leptogenesis, even though the
optimal value of the branching ratio depends in this region on the
specific values of the decay and annihilation rates.

\section{Discussion\label{sec_concl}}

Before we discuss our results, we remind the reader that most of the
conclusions drawn in the left-right symmetric seesaw model depend on
the fact that the eigenvalues of the Yukawa coupling matrix $y$
contain a large hierarchy. Some statements depend even on the fact
that we assume $y$ to be similar to the up-quark mass matrix. On the
other hand, we believe this to be the natural choice in left-right
symmetric models. In seesaw models of pure type I, a large hierarchy
in the Yukawa couplings requires a certain amount of fine-tuning,
since the Majorana coupling matrix $f$ inherits the doubled hierarchy
of $y$. In type I+II models, and particularly in the left-right
symmetric model under consideration, this does not hold true so that
these models can be more easily embedded into a GUT, what motivates
our choice for the Yukawa coupling matrix.

First of all, viable triplet leptogenesis requires that the triplet is
lighter than the lightest right-handed neutrino, otherwise inverse
decays erase every produced lepton asymmetry. In the present
left-right symmetric model, this implies that only the two solutions
for the Majorana masses of type '$\pm++$' can account for the observed
baryon asymmetry, and additionally $v_R/v_L \gtrsim 10^{22}$ is
required. Concerning the CP-asymmetry $\epsilon_L$, we found that the
upper bound presented in ref.~\cite{Hambye:2005tk} can be
approximately (up to a factor 2) saturated by utilizing the Majorana
phases in the Yukawa coupling matrix $y$. In addition, $\epsilon_L$ is
nearly constant for values $v_R/v_L < 10^{28}$. Hence, the prospects
of triplet leptogenesis are promising in the region
\be
10^{22} \lesssim v_R/v_L \lesssim 10^{28}.
\ee
Notice from Fig.~\ref{fig:mtildep001} that in this regime the
parameter $\tilde m_T$ varies between $m_0$ (the lightest left-handed
neutrino mass) and a few tenths of eV such that a judicious choice of
$v_R/v_L$ can be used to specify $\tilde m_T$ in this region, while
the parameter $\epsilon_L$ is unaffected. The remaining free
parameters in the left-right symmetric model can be used to adjust the
branching ratios $B_L$ and $B_H$. This implies that in the present
model the parameters $B_H$ ($B_L$), $\tilde m_T$, and $M_T$ can in
certain ranges be considered as independent parameters.

The results for the baryon asymmetry can be qualitatively understood
in different limits as follows. For large triplet decay rates, the
gauge interactions are irrelevant. In this case, washout is very
strong and leads to exponential suppression unless one of the
branching ratios is tiny. On the other hand, for small triplet decay
rates, inverse decays are not relevant such that the efficiency $\eta$
is almost independent of $B_L$ and $B_H$. In this regime, the produced
baryon asymmetry is, as the CP-asymmetry, proportional to $\sqrt{B_L
B_H}$, and hence, maximized by the choice $B_L=B_H=1/2$.  The most
promising region for leptogenesis is given be the intermediate regime,
$\gamma_D \approx \gamma_A$ at $z\approx 1$, and $B_L \approx B_H$,
which is shown in Fig.~\ref{eta_plot}. However, this region is not
accessible by the left-right symmetric model, since generically
$\tilde m_T \gtrsim 0.05$ eV, and the triplet decay rate exceeds the
annihilation rate for light triplets, $M_T \lesssim 10^{12}$ GeV. This
makes a hierarchy in the branching ratios necessary and we find the
lower bound on the triplet mass $M_T$ to be given by
\be
M_T \gtrsim 1.0 \times 10^{11} \textrm{ GeV } (\tilde m_T= 0.05 \textrm{ eV}).
\ee
With the optimal choice $B_H = 10^{-3}$, the produced baryon
asymmetry is then given by
\bea
\eta &=& 9.1 \times 10^{-2},  \\
\epsilon_{L,\textrm{max}} &=& 7.2 \times 10^{-7} \frac{m_0}{0.05 \textrm{ eV}},  \\
\eta_B &=& 25.7 \times 10^{-10} 
\frac{m_0}{0.05 \textrm{ eV}}
\frac{\epsilon_L}{\epsilon_{L,\textrm{max}}},
\eea
which is in accordance with observation for $m_0\approx 0.05$ eV and
for an almost maximal CP-asymmetry. This is slightly better than the
result obtained in ref.~\cite{Hambye:2005tk}, since we maximized the
baryon asymmetry with respect to $B_H$. Nevertheless, the prospects of
triplet leptogenesis are generally worse than the ones of leptogenesis
driven by right-handed neutrino decays in the left-right symmetric
model~\cite{Akhmedov:2006yp} and are not compatible with the bounds on
the reheating temperature coming from gravitino physics in
supersymmetric models.

\acknowledgments
This work was supported by the G\"oran Gustafsson Foundation [T.H. and
T.O.], the Swedish Research Council (Vetenskapsr{\aa}det), contract
nos. 621-2001-1611 [T.K. and T.O.] and 621-2005-3588 [T.O.], and the
Royal Swedish Academy of Sciences (KVA) [T.O.].

\bibliographystyle{apsrev} 

\bibliography{references} 

\end{document}